\documentclass[conference]{IEEEtran}
\IEEEoverridecommandlockouts
\pdfoutput=1
\usepackage{cite}
\usepackage{amsmath,amssymb,amsfonts}
\usepackage{algorithmic}
\usepackage{graphicx}
\usepackage{textcomp}
\usepackage{xcolor}
\usepackage[utf8]{inputenc}
\usepackage{mathtools,array}
\usepackage{etoolbox}
\usepackage{color}
\makeatletter
\patchcmd{\@makecaption}
  {\scshape}
  {}
  {}
  {}
\makeatother

\def\BibTeX{{\rm B\kern-.05em{\sc i\kern-.025em b}\kern-.08em
    T\kern-.1667em\lower.7ex\hbox{E}\kern-.125emX}}
\begin{document}
\title{H\&E Stain Normalization using U-Net\\
}

\author{
    \IEEEauthorblockN{Chi-Chen Lee\IEEEauthorrefmark{1}, Po-Tsun Paul Kuo\IEEEauthorrefmark{1}\IEEEauthorrefmark{2}, Chi-Han Peng\IEEEauthorrefmark{1}}
    
    \IEEEauthorblockA{\IEEEauthorrefmark{1}National Yang Ming Chiao Tung University (NYCU), Taiwan
    \\ \{ww888225433.ai10, pengchihan\}@nycu.edu.tw}
    
    \IEEEauthorblockA{\IEEEauthorrefmark{2}Advantech Co., Ltd., Taiwan
    \\ paul.kuo@advantech.com.tw}
}

\maketitle

\begin{abstract}
We propose a novel hematoxylin and eosin (H\&E) stain normalization method based on a modified U-Net neural network architecture. Unlike previous deep-learning methods that were often based on generative adversarial networks (GANs), we take a teacher-student approach and use paired datasets generated by a trained CycleGAN to train a U-Net to perform the stain normalization task.
Through experiments, we compared our method to two recent competing methods, CycleGAN and StainNet, a lightweight approach also based on the teacher-student model. We found that our method is faster and can process  larger images with better quality compared to CycleGAN. We also compared to StainNet and found that our method delivered quantitatively and qualitatively better results.


\end{abstract}

\section{Introduction}

Histopathology is the examination of human tissues under a microscope to study the manifestation of diseases. The tissues under study come from a biopsy or surgical procedure, then are processed and cut into very thin layers, and then stained and examined by pathologists under microscopes to characterize the details of the cells in the tissue. Computer-aided diagnosis (CAD) systems to assist the analysis are also common. Conventionally, the tissue is stained with hematoxylin and eosin (H\&E)~\cite{HE}. However, the stain images can have strong variances (in terms of colors, illumination, image qualities, etc.) due to differences in the image acquisition processes, such as tissue fixation duration, compositions of the H\&E-stains, or scanner settings~\cite{Sophia21}. The strong variances can hinder downstream tasks such as classification. In histopathology, {\em stain normalization} can mitigate the variances problem by transforming a stain image done by one kind of staining process to one done by another.

Deep learning (DL)-based methods for stain normalization are gaining popularity in recent years, showing fast improving performances when compared to traditional image processing-based approaches~\cite{Atefeh22, Sophia21}. In general, they conduct style-transfer tasks using neural networks such as Generative Adversarial Networks (GANs). Common choices for the network structures include: 1) cycle-consistent adversarial networks ("CycleGANs"~\cite{CycleGAN} in short), 2) GANs with disentangled feature presentations (e.g., DRIT++~\cite{DRIP++}), and 3) multi-task convolutional neural networks (CNNs)~\cite{Marini21}.

Although deep learning-based methods performed well in transferring the colors, there are still some limitations. GAN-based methods have complicated neural network structures. Therefore, they are more computationally expensive to train and inference. They also impose "patch consistency" problems when the input image is too large and needs to be partitioned into smaller patches before they can processed by neural networks. On the contrary, StainNet~\cite{StainNet} proposed a much simpler fully 1x1 convolutional network to transform colors in a pixel-to-pixel manner. StainNet's results are slightly less accurate than GAN-based methods (measured by GAN-related metrics such as FID (Fréchet Inception Distance)), but is much faster then CycleGAN-based approaches and has more consistent results across patches because this model learns pixel-wise property rather than the pixel distributions that were learned for CycleGAN.  


To address the shortcomings of existing methods, we propose a novel stain normalization method that is based on a lightweight {\em U-Net}~\cite{UNet} neural network architecture. We trained our network on a bidirectionally paired dataset produced by CycleGAN. Quantitative results show that our network works well on color transfers between two types of H\&E stains and can achieve better cycle consistency than the standard CycleGAN-based method. Besides, our output images obtain better FID scores than StainNet's results. Qualitative results show that our normalized images look as similar to real ones as CycleGAN's results. In contrast, StainNet's results have slight but notable color shifts. Same as StainNet, our results are more consistent across partitioned patches than CycleGAN's.

\section{Related Work}


\subsection{Image Processing-based Methods}

Even simply converting all stain images to grayscale may be beneficial to certain downstream tasks in medicine~\cite{Hamilton97, Ruiz07}. More modern approaches seek to retain the color information. Deconvolutional methods find a transform function between ground-truth source and target image pairs and use the function to transform unseen source images. Usually, the source and target images are first compressed into small "stain vectors" for the transform function to deduct from. These vectors are chosen manually~\cite{Ruifrok01} or by statistical algorithms such as Singular Value Decomposition (SVD)~\cite{Macenko09} or segmentation and clustering~\cite{SCAN}. Template color matching algorithms~\cite{Reinhard01} attempt to match distributions of each color channels between source and target images by various approaches such as variational Bayesian Gaussian mixture models (GMMs)~\cite{Magee09} and histogram normalization of the colormaps~\cite{Kothari11}.

\subsection{Deep Learning-based Methods}


Many stain normalization methods are based on generative adversarial networks (GANs). In short, GANs are suitable for conducting {\em image-to-image translation}~\cite{pix2pix} tasks in which images of one group (i.e., a "style") can be transformed into one of another group. Note that GAN-based style transfer is a kind of {\em unsupervised learning}, meaning that images in the two groups don't need to be paired. This is a major advantage over many traditional methods in which paired images are needed. StainGAN~\cite{StainGAN} trained a cycle-consistent adversarial network ("CycleGAN"~\cite{CycleGAN}) to transfer H\&E stain images from one scanner style to another (i.e., Hamamatsu Nanozoomer 2.0-HT to Aperio Scanscope XT). Similarly, Runz et al.~\cite{Runz21} investigated the potential and limitations of using CycleGAN for transfering H\&E stain images.~\cite{Sophia21} and~\cite{Atefeh22} used GANs with disentangled feature presentations (e.g., DRIT++~\cite{DRIP++}) to solve the stain normalization problem. Marini et al.~\cite{Marini21} proposed a novel convolutional neural network (CNN)-based approach that aims to tackle highly heterogeneous images. 

In StainNet~\cite{StainNet}, the authors proposed a lightweight fully 1x1 convolutional network to conduct color normalization in a pixel-to-pixel manner. Their method is much faster than GAN-based methods but performed slightly worse measured by GAN-related metrics such as FID (Fréchet Inception Distance) scores. They took a teacher-student model in which their neural network is trained on {\em paired} images synthesized by StainGAN. Our U-Net-based method is most similar to theirs.


\section{Method}

\subsection{Dataset}

The HE-Staining Variation (HEV) dataset~\cite{Runz21}, produced by the Institute of Pathology at Heidelberg University, was used to train and evaluate our model. The dataset offers serial sections of a follicular thyroid carcinoma, stained with different HE-staining protocols. We work on two styles of images. The first is done by a standard H\&E stain process (of the Institute of Pathology at Heidelberg University), denoted as "HE". The second are images intentionally stained too-long, denoted as "Long HE". Note that the dataset is {\em unpaired} - meaning that we don't have the exact Long HE counterpart of a HE image and vice versa. Each histopathological whole-slide image (WSI) is preprocessed into tiles of 256×256 pixels. We use 10,000 tiles of each style for training CycleGAN, 10,000 tiles of each style for inferencing CycleGAN to generate the training data for StainNet and U-Net, and 5000 HE image tiles for evaluating the performance of methods.

\subsection{Method Overview}

\begin{figure}[t]
  \centering
  \includegraphics[width=0.88\linewidth]{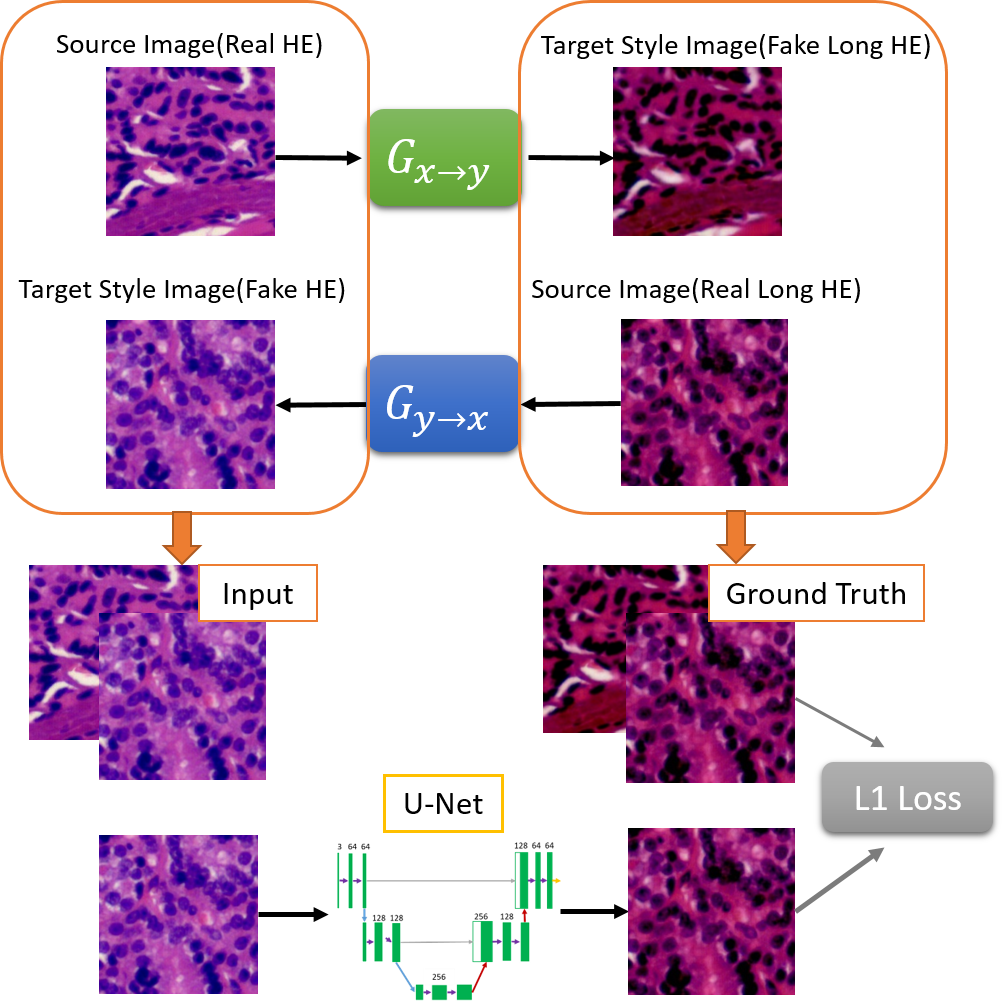}
  \caption{\label{fig:pipeline} Overview of our method. The $G_{x \rightarrow y}$ and $G_{y \rightarrow x}$ denote the generators of a trained CycleGAN that can synthesize a "fake" Long HE image from a real HE image and vice versa. We use the generated HE and Long HE image pairs, and the image pairs from the original HEV dataset (not shown here), to train a modified U-Net neural network that can be used to synthesize Long HE images from any HE images.}
\end{figure}

An overview of our method is shown in Fig.~\ref{fig:pipeline}. We first trained a CycleGAN using the HE and Long HE images in the HEV dataset. Next, inspired by the teacher-student model in deep learning, we used CycleGAN to generate synthetic (fake) Long HE images from real HE images and vice versa. In the end, we have paired HE and Long HE images to train a U-Net neural network. 
As shown in top of Fig.~\ref{fig:pipeline}, we also enlarged our HE to Long HE training dataset by combining real and fake HE images as input and combining real and fake Long HE images as ground truth to improve quantitative performance of our model.

\subsection{Neural Network Architecture}

\begin{figure}[t]
  \centering
  \includegraphics[width=0.78\linewidth]{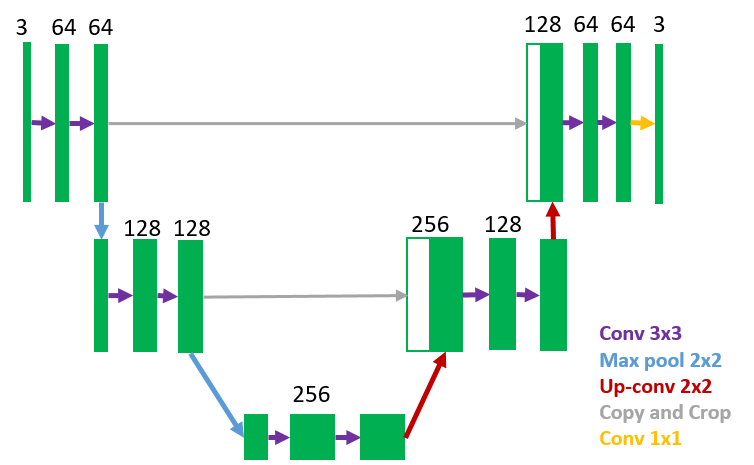}
  \caption{\label{fig:netwrok architecture} Our U-Net neural network architecture.}
\end{figure}

As shown in Fig.~\ref{fig:netwrok architecture}, our U-Net architecture is symmetric and  consists of a contraction and an expansion sections. The contraction section consists of two contraction blocks. Each contraction block takes an input and passes it to two 3x3 convolution layers followed by a 2x2 max pooling. We double the number of feature channels and halve the size of feature maps after each contraction block. The  expansion section also consists of two expansion blocks. Each block passes the input to two 3x3 CNN layers followed by a 2x2 up convolution layer that halves the number of feature channels. We combines the feature and spatial information through a sequence of up-convolutions and concatenations with high-resolution features from the corresponding contraction block. At the final layer a 1x1 convolution layer is used to map each 64-component feature vector to a 3-channel vector.

\subsection{Implementation}

We use CycleGAN to generate the training dataset for both U-Net and StainNet. The CycleGAN model was trained using Adam optimizer for 60 epochs. For U-Net, the model was trained with SGD optimizer for 300 epochs, and we use cosine annealing scheduler to decrease the learning rate from 0.01 to 0. The L1 loss function is used to minimize the error which is the average of the all the absolute differences between the output image and the ground truth. We use a NVIDIA RTX 3060 Ti machine.

\section{Results and Analysis}

\begin{table}[]
\begin{center}
\caption{Times (in seconds) to generate 5000 256x256 Long HE stain images using CycleGAN, StainNet, and our U-Net based method.}
\begin{tabular}{|c|c|c|c|}
\hline
        & CycleGAN & StainNet & Our  \\ \hline
Time(s) & 271.18   & 123.22   & 205.05 \\ \hline
\end{tabular}
\label{table:1}
\end{center}
\end{table}

We first compare the inference speeds of our method versus a CycleGAN-based stain normalization method and StainNet. As shown in Table~\ref{table:1}, our method is 24.39\% faster than CycleGAN, but is 66.4\% slower than StainNet.

To quantitatively evaluate our method, we measure the distances between distributions of our generated images and real images using Fréchet Inception Distance (FID), which is a standard way to evaluate the quality of results of generative models. In addition, to measure cycle consistency, we transfer HE stain images to the Long HE domain and transfer them back, and then calculate the similarity between each original HE image and its reconstructed one. Following the procedures in the previous papers~\cite{Runz21,StainGAN}, we use Structural Similarity Index Measure (SSIM) and Peak Signal to Noise Ratio (PSNR) to compare images. These metrics are explained as follows:


{\em \textbf{Fréchet Inception Distance (FID)}} is the squared Wasserstein distance between two multidimensional Gaussian distributions, $N(\mu,\Sigma)$ and $N(\mu_w,\Sigma_w)$:
\begin{equation*} 
\begin{aligned} 
FID = ||\mu - \mu_w||_2^2 + tr(\Sigma + \Sigma_w - 2(\Sigma^\frac{1}{2} \cdot \Sigma_w \cdot \Sigma^\frac{1}{2})^\frac{1}{2}),
\end{aligned}
\end{equation*}
where $\mu$ and $\Sigma$, and $\mu_w$ and $\Sigma_w$, are the mean vector and covariance matrix of some neural network internal representations of the generated images and real images, respectively. $tr$ is the trace of a matrix. The Inception v3~\cite{InceptionV3} trained on the ImageNet is commonly used as the internal representation.


{\em \textbf{Structural Similarity Index (SSIM)}} is used to measure the similarity between two images. It is the mean of the following score calculated over many windows, $x$ and $y$, of the given image pair:
\begin{equation*}
SSIM(x,y)=\frac{\left(2\mu_x\mu_y+c1\right)\left(\sigma_{xy}+c2\right)}
{\left(\mu_x^2+\mu_y^2+c1\right)\left(\sigma_x^2+\sigma_y^2+c2\right)},
\end{equation*}
where $\mu_x$ and $\mu_y $, and $\sigma_x^2$ and $\sigma_y^2$, are the average and variance of $x$ and $y$, respectively. $\sigma_{xy}$ is the covariance of $x$ and $y$. The range of SSIM is from 0 (least similar) to 1 (most similar).

{\em \textbf{Peak Signal to Noise Ratio(PSNR)}} also measures the similarity of an original image versus a generated image:
\begin{equation*}
PSNR = 20 log_{10} \frac{MAX_f}{\sqrt(MSE)},
\end{equation*}
where $MAX_f$ is the maximal value of the original image and $MSE$ is the Mean Squared Error of the two images. Higher PSNR value mean more similarity and vice versa.



\begin{table}[t]
\centering
\caption{FID scores between real Long HE images and generated Long HE images.}
\begin{tabular}{|c|c|c|c|}
\hline
    & CycleGAN & StainNet & Our \\ \hline
FID & 16.97    & 24.83    & 18.07 \\ \hline
\end{tabular}
\label{table:2} 
\end{table}

\begin{table}[t]
\centering
\caption{Cycle Consistency scores between real HE images and reconstructed HE images.}
\begin{tabular}{|c|c|c|c|}
\hline
      & CycleGAN & StainNet & Our  \\ \hline
SSIM  & 0.9554   & 0.9637   & 0.9672
  \\ \hline
PSNR  & 36.07    & 34.37    & 36.27  \\ \hline

\end{tabular}
\label{table:3}
\end{table}

\begin{figure*}[t]
  \centering
  \includegraphics[width=0.84\linewidth]{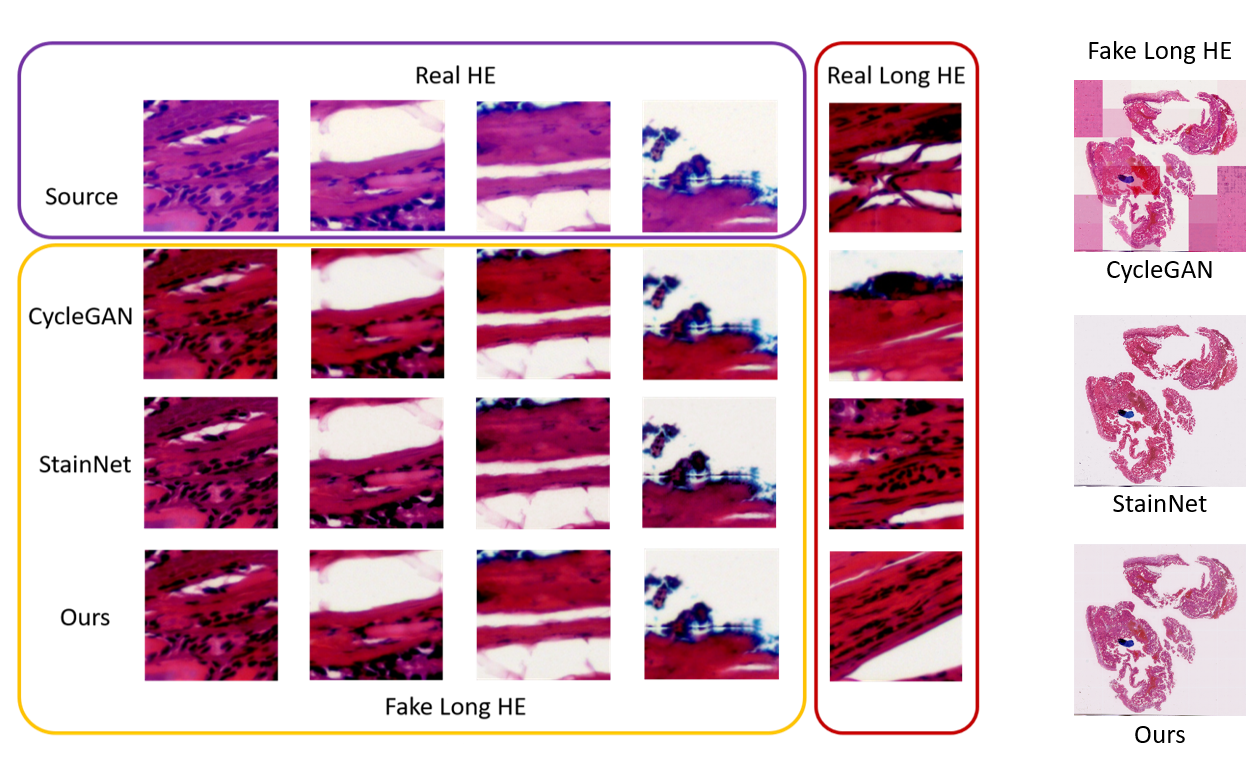}
  \caption{\label{fig:Results}  Left: Qualitative results of various stain normalization methods, which transfer from HE to Long HE. The right column shows some real Long HE. Right: large inputs (3120x3168) are partitioned into smaller (512x512) patches before they can be processed by the stain normalization methods.} 
\end{figure*}

In Table~\ref{table:2}, we show the FID scores between real Long HE images and fake Long HE images generated by the three methods. As expected, our method's FID score is slightly worse (higher) than CycleGAN's because our method is trained on the data generated by CycleGAN. However, our method has significantly better FID score than StainNet (also a teacher-student model), meaning that our results are statistically more similar to ground truth. In Table~\ref{table:3}, we show cycle consistency scores of the three methods. Our method outperforms other two methods in terms of SSIM and PSNR scores, which shows that our network retains the source image information better. We illustrate  qualitative comparison of the methods in Fig.~\ref{fig:Results} left. We can see that fake Long HE images generated by CycleGAN and our method are more visually similar to real ones than StainNet's results (which have some color shifts). In Fig.~\ref{fig:Results} right, we show that our method, similar to StainNet, maintains better consistency between different partitioned patches of the input image than CycleGAN.


\section{Conclusion}

In this paper, we explore the novel idea of using U-Net to perform H\&E stain normalization tasks. Through experiments, we show that our method is significantly faster than a competing CycleGAN-based approach while still delivering similar results measured quantitatively and qualitatively. Compared to StainNet, our method is slower but delivers results with better qualities. For future work, we aim to improve the speed of our method while still maintaining the qualities and investigate the improvement of the downstream tasks (such as cancer cell detection) by this faster stain normalization towards larger digital pathological images.

\section{Acknowledgements}

This work is funded by the Ministry of Science and Technology of Taiwan (110N007). Chi-Chen Lee is a research intern at Advantech Co., Ltd., Taiwan.








\end{document}